\documentclass[12pt]{article}

\begin{document}

%%%%%%%%%%%%%%%%%%%%%%%%%%
% Vatche Sahakian's macros

\newcommand{\bb}{\begin{equation}}
\newcommand{\ee}{\end{equation}}
\newcommand{\bbb}{\begin{eqnarray}}
\newcommand{\eee}{\end{eqnarray}}
\newcommand{\vc}[1]{\mbox{$\vec{{\bf #1}}$}}
\newcommand{\mc}[1]{\mathcal{#1}}
\newcommand{\del}{\partial}
\newcommand{\lk}{\left}
\newcommand{\ave}[1]{\mbox{$\langle{#1}\rangle$}}
\newcommand{\re}{\right}
\newcommand{\pd}[1]{\frac{\del}{\del #1}}
\newcommand{\pdd}[2]{\frac{\del^2}{\del #1 \del #2}}
\newcommand{\Dd}[1]{\frac{d}{d #1}}
\newcommand{\sech}{\mbox{sech}}
\newcommand{\pref}[1]{(\ref{#1})}

\newcommand
{\sect}[1]{\vspace{20pt}{\LARGE}\noindent
{\bf #1:}}
\newcommand
{\subsect}[1]{\vspace{20pt}\hspace*{10pt}{\Large{$\bullet$}}\mbox{ }
{\bf #1}}
\newcommand
{\subsubsect}[1]{\hspace*{20pt}{\large{$\bullet$}}\mbox{ }
{\bf #1}}

\def\ie{{\it i.e.}}
\def\eg{{\it e.g.}}
\def\cf{{\it c.f.}}
\def\etal{{\it et.al.}}
\def\etc{{\it etc.}}

\def\AA{{\cal A}}
\def\BB{{\cal B}}
\def\CC{{\cal C}}
\def\DD{{\cal D}}
\def\EE{{\cal E}}
\def\FF{{\cal F}}
\def\GG{{\cal G}}
\def\HH{{\cal H}}
\def\II{{\cal I}}
\def\JJ{{\cal J}}
\def\KK{{\cal K}}
\def\LL{{\cal L}}
\def\MM{{\cal M}}
\def\NN{{\cal N}}
\def\OO{{\cal O}}
\def\PP{{\cal P}}
\def\QQ{{\cal Q}}
\def\RR{{\cal R}}
\def\SS{{\cal S}}
\def\TT{{\cal T}}
\def\UU{{\cal U}}
\def\VV{{\cal V}}
\def\WW{{\cal W}}
\def\XX{{\cal X}}
\def\YY{{\cal Y}}
\def\ZZ{{\cal Z}}

\def\sinh{{\rm sinh}}
\def\cosh{{\rm cosh}}
\def\tanh{{\rm tanh}}
\def\sgn{{\rm sgn}}
\def\det{{\rm det}}
\def\trace{{\rm Tr}}
\def\exp{{\rm exp}}
\def\sh{{\rm sh}}
\def\ch{{\rm ch}}

\def\ell{{\it l}}
\def\str{{\it str}}
\def\lp{\ell_{{\rm pl}}}
\def\blp{\overline{\ell}_{{\rm pl}}}
\def\ls{\ell_{{\str}}}
\def\bls{{\bar\ell}_{{\str}}}
\def\bM{{\overline{\rm M}}}
\def\gs{g_\str}
\def\gym{{g_{Y}}}
\def\geff{g_{\rm eff}}
\def\eff{{\rm eff}}
\def\r11{R_{11}}
\def\kel{{2\kappa_{11}^2}}
\def\kten{{2\kappa_{10}^2}}
\def\lpten{{\lp^{(10)}}}
\def\alp{{{\alpha}'}}
\def\alpe{{{\alpha_e}}}
\def\le{{{l}_e}}
\def\aleff{{\alp_{eff}}}
\def\sqaleff{{\alp_{eff}^2}}
\def\tgs{{\tilde{g}_s}}
\def\talp{{{\tilde{\alpha}}'}}
\def\tlp{{\tilde{\ell}_{{\rm pl}}}}
\def\tr11{{\tilde{R}_{11}}}
\def\wtilde{\widetilde}
\def\what{\widehat}
\def\hlp{{\hat{\ell}_{{\rm pl}}}}
\def\hr11{{\hat{R}_{11}}}
\def\hf{{\textstyle\frac12}}
\def\coeff#1#2{{\textstyle{#1\over#2}}}
\def\CY{Calabi-Yau}
\def\lessapprox{\;{\buildrel{<}\over{\scriptstyle\sim}}\;}
\def\greaterapprox{\;{\buildrel{>}\over{\scriptstyle\sim}}\;}
\def\inbar{\,\vrule height1.5ex width.4pt depth0pt}
\def\IC{\relax\hbox{$\inbar\kern-.3em{\rm C}$}}
\def\IR{\relax{\rm I\kern-.18em R}}
\def\IP{\relax{\rm I\kern-.18em P}}
\def\Z{{\bf Z}}
\def\R{{\bf R}}
\def\One{{1\hskip -3pt {\rm l}}}
\def\sst{\scriptscriptstyle}
\def\osc{{\rm\sst osc}}
\def\lam{\lambda}
\def\lc{{\sst LC}}
\def\pr{{\sst \rm pr}}
\def\cl{{\sst \rm cl}}
\def\D{{\sst D}}
\def\bh{{\sst BH}}
\def\vev#1{\langle#1\rangle}

\input{epsf}

\begin{titlepage}
\rightline{CLNS 00/1684}

\rightline{hep-th/0008073}

\vskip 2cm
\begin{center}
\Large{{\bf 
The phases of 2D NCOS
}}
\end{center}

\vskip 2cm
\begin{center}
Vatche Sahakian\footnote{\texttt{vvs@mail.lns.cornell.edu}}
\end{center}
\vskip 12pt
\centerline{\sl Laboratory of Nuclear Studies}
\centerline{\sl Cornell University}
\centerline{\sl Ithaca, NY 14853, USA}

\vskip 2cm

\begin{abstract}

We study the phases of the 1+1 dimensional Non-Commutative 
Open String theory on a circle. We find that 
the length scale of non-commutativity increases at strong coupling, the
coupling in turn being dressed by a power of D-string charge. 
The system is stringy at around this length scale, with 
dynamics involving an interplay between the open and 
wrapped closed strings sectors.
Above this energy scale and at strong coupling, and below it
at weak coupling, the system acquires a less stringy character.
The near horizon geometry of the configuration exhibits several
intriguing features, such as a flip in the dilaton field and the curvature
scale, reflecting UV-IR mixing in non-commutative dynamics. 
Two special points in the parameter measuring the size of
the circle are also identified.

\end{abstract}

\end{titlepage}
\newpage
\setcounter{page}{1}

\section{Introduction and discussion}

It has long been suspected that non-commutativity of space-time coordinates 
is a key ingredient in the proper
formulation of Planck scale physics~\cite{DKPS,MAT1}. 
In the context of string theory, this
phenomenon has risen as a characteristic feature of strong coupling 
dynamics.
A common theme in exploring new string physics has been
to arrange for systems where an unknown regime of the theory can be studied
through a dual formulation.
The Holographic duality~\cite{MALDA1} for example
associates a low energy sector of certain
open string theories with supergravity, leading to a better understanding
of Super Yang-Mills (SYM) theories at strong coupling. 
In this setup however,
either too much of the interesting string dynamics is
scaled away, so that one essentially learns about field theories;
or the stringy remnant in the decoupled regime is not very well understood,
as in the scenario of little string theories. 
It would naturally be 
desirable to find settings where more of stringy dynamics, such as
non-commutativity of spacetime coordinates, survives
the decoupling limit. Indeed, in the work of~\cite{SWNC}, it has been
shown that this may be achieved by adding another charge to 
certain D brane systems.
One is then lead to a spectrum of 
new string/field 
theories on non-commutative spaces~\cite{SWNC}-\cite{KAWASASA}. 
And through
the Holographic duality, even the strong coupling regimes of these
theories can now be explored~\cite{HASHITZNC,MALDARUSSO}. 

In this context, a particularly interesting setup is obtained
by adding fundamental string charge to a system of D-strings. 
The resulting  bound state of strings and D-strings can be studied
through a 1+1 dimensional Non-Commutative Open String theory (NCOS)
~\cite{GMSS,KLEBMALDA}.
This system appears to be the simplest one that explores non-commutative
dynamics; its role in understanding this phenomenon may be as fundamental
as that of the role of D0 branes in understanding Dp brane dynamics~\cite{MAT1}. 
The theory, unlike little string theories and OM theory, has a well defined
perturbative expansion.
It also necessarily involves non-commutativity in the time coordinate;
all this within the framework of a 
self-consistent, Lorentz invariant, kinematically simple
and computationally accessible string theory. It is then a good
candidate for a framework to put one's intuition with regards to
non-commutativity to extreme tests. Finally, the particularities of
non-commutativity involving the time variable may yield new
insight in understanding information transfer and dynamics
near black hole horizons.

In this work, we attempt to understand this theory in a thermodynamic
setting, in the hope that such a macroscopic analysis would be a guide
to identifying some of the new interesting dynamical consequences of 
time-space non-commutativity. We expect that critical phenomena in
the system would reflect the peculiarities of the underlying
microscopic dynamics.

Most of the technical aspects of our analysis will
parallel similar ones that have appeared in the literature~\cite{MSSYM123} 
in the context of commutative Yang-Mills theories. In the next section, 
we briefly review the theory of interest.
In Section~\ref{phasediag},
we present the phase diagram of 1+1 dimensional NCOS theory on a 
circle, with
a discussion of the physics underlying the various phases.
In Section~\ref{disc}, we summarize the new results of the paper, and
suggest future directions. Section~\ref{details} contains all the
details of the construction of the phase diagram and may be skipped
without much grief to anyone. Finally, the Appendix contains 
a roadmap used extensively in the text of Section~\ref{details}.

\vspace{0.5cm}
{\bf Note added:} The article~\cite{HARMARK} also studies this system,
as well as the higher dimensional cases, and reaches some of the same 
conclusions.

\subsection{Preliminaries}\label{prem}

In this section, we define the 
1+1 dimensional NCOS theory~\cite{GMSS,KLEBMALDA} and set
the notation and conventions used in the paper. We start in IIB theory
in the presence of a bound state of
N fundamental strings and M D-strings, wrapped on a coordinate $y$ 
of size $\Sigma$. 
The theory is 
parameterized by the string coupling $\gs$ and string scale $\alp$.
We choose coordinates such that the metric is given by
\bb\label{strangemetric}
g_{\mu\nu}=\gs \eta_{\mu\nu}\ .
\ee
The NSNS B field then becomes
\bb\label{quant}
B^2\equiv\frac{\lk(2\pi\alp\re)^2 B_{ty}^2}{\gs^2}=
\lk(1+\lk(\frac{M}{N\gs}\re)^2 \re)^{-1}\ .
\ee
The dynamics of this bound state can be described by
a 1+1 dimensional theory of open strings propagating on a 
non-commutative space with metric
\bb
G_{\mu\nu}=\gs \lk(1-B^2\re)\eta_{\mu\nu}\ .
\ee
The string endpoints carry M Chan-Paton indices and
are confined to the non-commutative $t-y$ plane
\bb\label{noncomm}
[t,y]=i\theta\ ,
\ee
with the parameter $\theta$ given by
\bb
\theta=\frac{B^2}{B_{ty} \lk(1-B^2\re)}\ .
\ee
The open string coupling $G_o$ becomes
\bb
G_o^2=G_s=\gs \sqrt{1-B^2}\ .
\ee
It was argued in~\cite{SWNC,GMSS,KLEBMALDA} 
that there exists an energy regime where
the dynamics of these open strings decouples from the closed string
sector in the bulk. This decoupling limit is obtained by
\bb
\alp\rightarrow 0\mbox{        while keeping $\gs\alp$ fixed.}
\ee
The NSNS B field (or equivalently the electric field
on the D strings) then attains its maximal value
\bb
1-B^2\rightarrow \lk(\frac{M}{N\gs}\re)^2\rightarrow 0\ .
\ee
The metric becomes
\bb
G_{\mu\nu} \rightarrow \gs^{-1} \lk(\frac{M}{N}\re)^2\eta_{\mu\nu}\ ,
\ee
with $\theta$ given by
\bb
\theta\rightarrow 2\pi \gs\alp \lk(\frac{N}{M}\re)^2\equiv 2\pi\alpe\ .
\ee
The NCOS string scale $\alpe=\le^2$ is the fundamental length scale of 
this new string theory. While
the open string coupling becomes a rational number
\bb
G_s\rightarrow \frac{M}{N}\ .
\ee
This limit is survived by
the Virasoro tower of open string excitations since
\bb
G^{tt} E^2\sim \frac{N_{osc}}{\alp}\Rightarrow E^2\sim \frac{N_{osc}}{\alpe}\ ;
\ee
along with closed strings that wrap the cycle $y$.
The box size seen by the NCOS theory is $\Sigma$ and can be determined
by looking at the dispersion relation
of a wrapped closed string~\cite{KLEBMALDA}
\bb\label{wrapped}
\alp^2 E^2=\lk(\gs \omega \Sigma\re)^2 \lk(1-B^2\re)
-2 E B \omega \Sigma \gs \alp+2\gs \alp N_{osc}
\Rightarrow E= \frac{\omega \Sigma}{2 \alpe}
+\frac{N_{osc}}{\omega \Sigma}\ ,
\ee
where $\omega$ is the winding number.
In the S-dual frame, $\gs\rightarrow 1/\gs$ and $g_{\mu\nu}\rightarrow
g_{\mu\nu}/\gs=\eta_{\mu\nu}$. Then the NSNS field scales as
$B^2\rightarrow 1/\gs\rightarrow 0$, while the dual non-commutative
parameter scales to zero as well $\theta\rightarrow \alp/\sqrt{\gs}
\rightarrow 0$. The S-dual theory is decoupled 1+1 dimensional
U(N) SYM theory with $M$ units of electric flux~\cite{GMSS,KLEBMALDA}.

\subsection{The phase diagram}\label{phasediag}

The theory of interest is 1+1 dimensional NCOS theory as defined in~\cite{GMSS}
and summarized in the previous section.
Alternatively, we are studying the phase diagram of 1+1 dimensional U(N)
SYM theory with electric flux on a compact cycle.
The NCOS parameter space consists of the NCOS string scale $\le$, the 
string coupling $G_o$, the size of the compact cycle $\Sigma$, and the
integer $M$ counting the number of D-strings. On the SYM side,
the parameters are the dimensionful Yang-Mills coupling $\gym^2$,
the size of the circle $\Sigma$, the rank 
of the gauge group $N$, and
the $M$ units of electric flux. This entire setup can also
be embedded in Light-Cone M theory.
We can then look at every part of the phase diagram of this system 
from these three different viewpoints.

\begin{figure}
\epsfxsize=13.5cm \centerline{\leavevmode \epsfbox{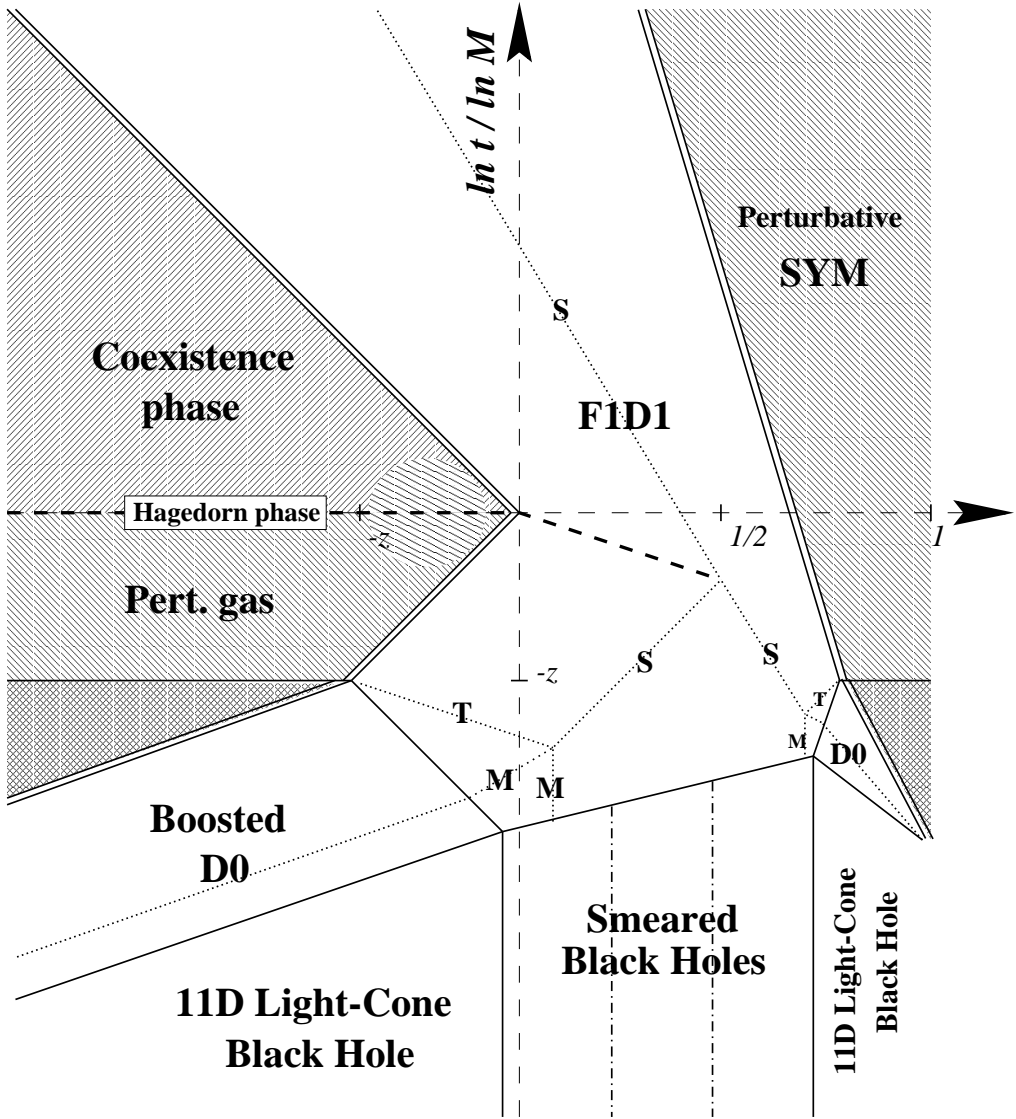}}
\caption{\sl The phase diagram for the 1+1 dimensional NCOS theory.
We have defined $z\equiv \ln \sigma /\ln M$. Solid lines are 
Gregory-LaFlamme phase transitions; double lines are string-scale
curvature regimes; dotted lines are duality transformations; shaded
regions have no valid supergravity duals. The horizontal axis
is the coupling, $\ln g/\ln M$.
}
\label{fig1}
\end{figure}

The scale of time-space non-commutativity of the NCOS theory
is set by $\le$. We fix $M$ and
$\sigma=\Sigma/\le$, and vary the coupling $g=G_o\sqrt{M}$ 
and the temperature $t=T\le$. These are
our four independent parameters. In Figure~\ref{fig1},
we plot $\ln t/\ln M$
on the vertical axis, and $\ln g / \ln M$ on the horizontal.
Varying $g$ corresponds to changing $N$ with fixed $M$;
this means that $g$ is restricted to the range
\bb
0\leq g \leq M\ ,
\ee
for $N>1$. Hence, the diagram is truncated on the right.
The set of available couplings is discrete. We will think of $M$
as being a large integer and think of the set as being dense
in the region of interest. 
This setup is most convenient from the perspective of the
NCOS theory. In the S-dual SYM theory, it corresponds to 
exploring a space of U(N) SYM theories with different ranks for
the gauge group; every vertical line on the diagram corresponds to a
different SYM theory as a function of energy scale.
We also choose the cycle size $\sigma$ in the range $1 \leq \sigma \leq M$,
keeping track of
it through the variable $z\equiv \ln \sigma /\ln M$, with
$0\leq z\leq 1$.

The global structure
of the diagram looks familiar from similar analysis applied to SYM
theories without electric flux~\cite{MSSYM123}. 
However, several new twists arise, 
and differences emerge that explore naturally the peculiarities of 
time-space non-commutativity. Before dwelling into details, let us
make some general comments regarding these differences.
Phases with different equations of state appear delineated
in Figure~\ref{fig1} by solid lines. Dotted lines correspond to duality
transformations that patch together different supergravity vacua 
within a given phase.
Shaded regions cannot be described by dual supergravity solutions.
Single lines correspond to 
Gregory-LaFlamme transitions~\cite{LAFLAMME1, LAFLAMME2}, while 
double lines are associated with string scale curvatures in the
corresponding geometries.
On the right, and at high temperatures, 
the structure of the diagram is similar to the one
arising in the case of zero electric flux. 
On the left, at temperatures
around the NCOS string scale, NCOS stringy dynamics sets in. In the middle,
at strong NCOS coupling $g$,
the scale of non-commutativity gets dressed by a power of the coupling and
the phase structure `folds' about a new energy scale
\bb\label{newnc}
t\sim g^{-1/3}\equiv t_c\ .
\ee
The scale of non-commutativity as a function of $g$
is shown in the figure by the dashed line.
A similar phenomena was noted in~\cite{HASHITZNC} in the context of D3 branes. 
For the most part, well below and above
the dashed line, the system may be described with commutative,
field theoretical dynamics, albeit a complicated one. We next
describe the dynamical details underlying the various phases, 
moving from the left towards the right.

The 1+1 dimensional NCOS theory on the circle
consists of three sectors: decoupled massless modes, corresponding
to the U(1) modes of the S-dual SYM theory; 
open strings on a non-commutative space;
and a sector of closed strings wrapping
the compact cycle. The $U(1)$ modes are irrelevant to the
thermodynamics, as they describe the dynamics of the center of mass of
the system. At $t\sim 1$, near the NCOS string scale, in the middle and
left of the phase diagram, we have a few long open strings at high
oscillator level, at the Hagedorn transition point. As we lower the
temperature at weak coupling, it becomes thermodynamically more
favorable to distribute the energy amongst many open strings, each at low
oscillator number. In the presence of non-commutativity, 
as a result of~\pref{noncomm}, these low lying modes have longitudinal 
extent $\Delta y$ proportional to the temperature~\cite{SST,YONEYA}
\bb\label{unc1}
\Delta y \sim t\ \le\ .
\ee
This is termed the non-commutative UV-IR relation.
For $t\ll 1$, this $\Delta y$ is substringy. On the other hand, the Compton
wavelength of these constituents grows with lower temperatures as~\cite{YONEYA}
\bb\label{unc2}
\Delta y \sim \le/t\ .
\ee
Therefore, the latter is
the relation that sets their characteristic size for $t\ll 1$. 
At $t\sim 1/\sigma$, we then expect finite size effects to set in.
Between $1/\sigma\ll t\ll 1$, we describe the
phase by a gas of weakly interacting massless
particles
\footnote{In an earlier version of the preprint, we had
incorrectly stated that the dynamics is in the center of mass
motion of massive low
oscillator number modes; as we see from the equation of state
below, the degrees of freedom are massless modes.  These 
excitations do not decouple from the massive NCOS open 
strings.}.
Given $M$ D-strings with large $M$, 
there are $M^2$ species of these animals. The equation
of state scales at leading order in the coupling  as
\bb\label{ncosone}
E\le \sim M^2 \sigma t^2\ .
\ee
As we move towards the right, there is a phase described by the
near horizon geometry of the bound state of $N$ fundamental strings
and $M$ D-strings, labeled in~Figure~\ref{fig1} as F1D1. 
Its equation of state is given by
\bb\label{ncosstrong}
E \le \sim \sigma M^2 g^{-1} t^3\ ,
\ee
reflecting strong coupling dynamics in the NCOS theory.
The
curvature of this geometry as measured at the horizon becomes string scale
at
\bb\label{tg}
t\sim g\ .
\ee
Equating~\pref{ncosone} with~\pref{ncosstrong}, we find~\pref{tg},
confirming the scenario just depicted.

The transition point $t\sim 1/\sigma$
sews onto a Gregory-LaFlamme transition at strong coupling to the
right, further
supporting the proposal to describe the phase above with a weakly
interacting gas of point-like particles. 
At temperatures below $t\sim 1/\sigma$, the phase diagram
mirrors structurally the right side. 
The region just below, shaded in dark in Figure~\ref{fig1}, 
is a transition phase whose dynamics has been 
a mystery even in the zero electric flux case~\cite{MSSYM123}. 
Further below, we reach a phase that can be
described by a supergravity vacuum, that of boosted D0 branes localized
along the compact cycle. The transition point is where the curvature
of the geometry as measured at the horizon becomes string scale.
For temperatures even lower, we connect to phases in Light-Cone
M theory, the Matrix theory realization of this setup. We will come 
to this regime later.

Coming back to the $t\sim 1$ point, we
increase the temperature, at fixed coupling, away
from the Hagedorn phase where the thermodynamics is described
by a few long NCOS open strings. Evidence was presented in~\cite{KLEBMALDA}
that the sector of wrapped closed strings of the NCOS theory may play
an important role in trusting the system past its Hagedorn `limiting'
temperature. Indeed, we can easily see this phenomenon here:
the equation of state of such closed strings
can be read off~\pref{wrapped}, and has the form
\bb\label{ncos2}
E\le\sim \omega \sigma t^2\ ,
\ee
where $\omega$ is the number of windings on the circle. 
The maximum value for $\omega$ is $N$, the number of D-strings
in the S-dual picture. Therefore, as we increase the
temperature from $t\sim 1$, the system may distribute its free energy
amongst excited closed strings that split off the NCOS open strings, 
gradually reaching the energy~\pref{ncos2} with $\omega=N$. In fact,
from the side of the supergravity dual, we find that the curvature
scale at the horizon at these high temperatures becomes string
scale around
\bb\label{tginv}
t\sim 1/g\ .
\ee
This is precisely the temperature scale at which equation~\pref{ncos2},
with $\omega=N$,
equals the equation of state at strong coupling~\pref{ncosstrong}, confirming
the scenario described. The phase between $1\ll t \ll 1/g$ is then 
described by a coexistence phase of open and closed strings. The 
existence of such a coexistence region also arose
in the context of the commutative SYM theory where
the analogous phase is that of the Matrix string~\cite{LMS,MSSYM123};
there, it was proposed
that such a transition region is needed to `unwind' the $Z_N$ holonomy
from the Matrix string configuration~\cite{DVV,MOTL}. In the current scenario,
the presence of D0 brane charge in the Light-Cone IIA theory in
question may change the dynamics.
But at these high temperatures,
equation~\pref{ncos2} with $\omega=N$ is the equation of state
of a IIA Matrix string with N units of Light-Cone momentum
\bb
E\sim \frac{\alp}{N\Sigma} \frac{S^2}{\alp}\ .
\ee
However, as we shall see below, 
this phase sews onto very different physics at
lower temperatures.

In passing, let us note that
both transition curves appearing in equations~\pref{tg} and~\pref{tginv}
relate the temperature to the open string coupling $G_o$ in the
combination $g=G_o\sqrt{M}$; $M$ does not appear in these equations
independently.
This is a motivation for identifying the relevant
coupling of the perturbative expansion of the theory as $g$ and not $G_o$.

\begin{figure}
\epsfxsize=12cm \centerline{\leavevmode \epsfbox{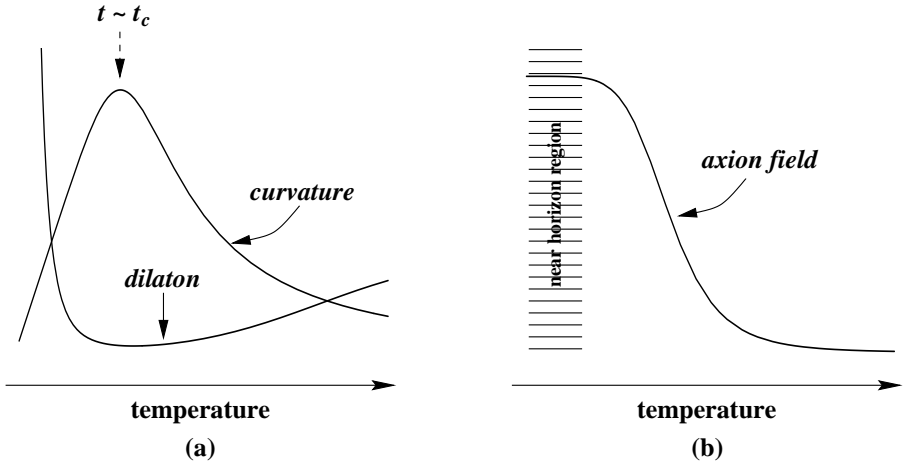}}
\caption{\sl (a) The curvature scale and coupling in the (N,M)
geometry in the near horizon region. (b) The axion field in
the same geometry; the attractor value is $N/M$.  } 
\label{fig2}
\end{figure}

We now move to the middle section of the phase diagram, where the dynamics
can be described by supergravity duals. The dominant patch in the
middle phase
consists of the near horizon geometry of the (N,M) string. The feature of
interest here is the emergence of a new energy scale for
non-commutativity given by equation~\pref{newnc}. The curvature at
the horizon and the string coupling are plotted in Figure~\ref{fig2}(a).
Both exhibit an extremum at the scale $t_c$. At weaker couplings towards the
left of the diagram, 
the effect of the flip in the curvature scale was elaborated on
above; it corresponds to the transition of dynamics between the
NCOS open and closed string sectors. This aspect does not
effect the Holographic UV-IR relation: we may have expected that the
dispersion relation in this background geometry is such that perturbations
seek areas of smaller curvature for lower energies; and this is
indeed the case. However, there also
exists a compensating red shift factor dressing the time at infinity, and
hence energy appears to still flow to lower scales
for lower values of the radial coordinate of the geometry. 

The flip in the dilaton flow at fixed NCOS coupling must reflect
the effects of the spreading of the degrees of freedom in the
longitudinal direction as a function of energy. Earlier, 
we argued qualitatively that we may expect that the constituents of the system
will spread in size 
as we move to lower and higher energies from the scale of non-commutativity
because of the quantum mechanical and stringy uncertainty relations,
respectively (see equations~\pref{unc2} and~\pref{unc1}). 
Overlapping wavefunctions would then imply stronger 
effective coupling between the degrees of freedom for energy scales away
from $t_c$. This then explains the peculiar behavior of the dilaton field
\footnote{We are assuming here that the flow of the dilaton is correlated with
the flow of the effective coupling of the Holographic dual degrees
of freedom.}. A phenomena of identical dynamics arose in the
context of D3 branes and T-duality in~\cite{HASHITZMOR}.
For high temperatures, we sew onto the S-dual geometry
of (-M,N) strings. For lower temperatures, we need to make use of 
another element of the SL(2,Z) duality group. A similar complication 
was also present in~\cite{HASHITZMOR} in the context of Morita
equivalence. The problem arises here since,
for $t\ll t_c$, the axion field that vanishes at infinity
is attracted to a fixed value in the near horizon region, complicating
the action of the S-duality group. This value of the axion field is
a rational $N/M$; on the other side of the duality, 
it is the inverse. 
Figure~\ref{fig2}(b) shows the behavior of the field
in this vacuum. Note also that the geometry for
$t\ll t_c$ is identical (up to a coordinate transformation)
to the one appearing on the right of the diagram
in the S-dual SYM field theory frame, further supporting the notion that at
low temperatures the non-commutative (hence stringy) 
aspects of the NCOS dynamics become unimportant.

As we move to lower temperatures in the middle of Figure~\ref{fig1},
a T-duality on $\Sigma$ is required leading us
to a phase of smeared boosted D0 branes; and then
a Gregory-LaFlamme transition on $\Sigma$ settles the system into
$M$ localized D0 branes with $N$ units of boost.
Further down, we lift to M theory, first to an oblique wave
on the torus, then to a boosted black hole. For these low energies, we
connect to a generic finite temperature vacuum in Light-Cone
M theory, suggesting that the NCOS theory can describe Light-Cone
M theory with an additional boost~\cite{GMSS}. 
We will elaborate on this issue further
down.

To the right of the diagram, at high temperatures, 
the strong coupling behavior of the NCOS theory
can be described by weakly coupled 1+1 dimensional
SYM degrees of freedom. The Yang-Mills coupling is given by
\bb
\gym^2=\frac{M^2}{g^4 \alpe}=\lk(\frac{N}{M}\re)^2 \alpe^{-1}\ ,
\ee
so that the coupling becomes a rational number when
measured at the NCOS string scale. The phase
structure in the right half of the diagram is virtually identical to the one
that arises in the 1+1 dimensional SYM case with zero electric flux.
We will therefore only briefly
discuss this part
and refer the reader to~\cite{MSSYM123}. We note that, in the current scenario,
each different vertical line on our diagram corresponds to a different
U(N) SYM theory, since $N$ is varying with $M$ being fixed. But
each vertical line maps onto a SYM theory with a shifted
Yang-Mills coupling, such that the effective Yang-Mills coupling
measured at a fixed temperature is increasing when one moves
toward the left. This coupling 
$\geff^2=\gym^2 N T^{-2}$, $T$ being the temperature, becomes
of order one when measured at the double line between the perturbative SYM gas 
and the F1D1 phases
\footnote{To structurally 
relate to Figure 1 of~\cite{MSSYM123}, we can identify the entropy there with
our vertical axis here, while the horizontal axis there is the
Yang-Mills coupling measured at the IR cutoff $\Sigma$; in our case,
this corresponds to the combination $M^4 \sigma^2/g^6$. The diagram
in~\cite{MSSYM123} 
does not truncate on the right since one varies the relative scales
between the Yang-Mills coupling and the box size $\Sigma$, \ie\ $\sigma$
in the current language, while keep the rank of the gauge group $N$ fixed.}.
The equation of state of the perturbative phase is given by~\pref{ncos2}
with $\omega=N^2$. More generally,
the $M$ units of electric flux in the SYM theory appear not to effect
equations of state or critical phenomena arising in this part of
the phase diagram. 
The region that appears checkered between the SYM gas and D0 phases 
involves a transition phenomenon that organizes the SYM excitations
into its zero modes at fixed entropy and large coupling. 
It appears in~\cite{MSSYM123} as a 
single horizontal solid line. At lower temperatures, we dwell into
phases of Light-Cone M theory, the Matrix theory regime.

Let us next focus on the phases appearing at the lowest temperatures.
As is typical in these systems, the preferred configurations are
eleven dimensional black holes in Light-Cone M theory.
On the left and right sides, these black holes are localized
and carry oblique momentum on a two torus. Both momenta survive the
decoupling limit; but the dominant one sets the Light-Cone direction.
The Planck scale on both sides is given in terms of the NCOS parameters by
\bb\label{lp}
\lp^3=\frac{\alp^3}{\le^3} \frac{M^2}{\sigma g^4}\rightarrow 0\ .
\ee
On the left side, the
eleventh cycle and the cycle $R$ related to $\Sigma$ scale as
\bb\label{radii}
\frac{\lp^2}{R}= \sigma^{1/3} M^{-2/3} g^{4/3} \le\ \ \ \ ,
\frac{\lp}{R_{11}}= \sigma^{2/3} M^{2/3} g^{-4/3}\ ,
\ee
which are held fixed in the decoupling limit. We see that the
Light-Cone circle is $R$, not $R_{11}$. On the right side, these
relations are the same with $R\leftrightarrow R_{11}$, reflecting an
action of the modular group of the torus in between. Hence, on the right side,
the Light-Cone direction is the eleventh direction. Correspondingly,
for both sides, the momentum in the Light-Cone direction is N units,
the dominant momentum charge in this decoupling limit.
The equation of state of the phase on the left is
\bb\label{bbheos}
E\le\sim g^{-6/7} \sigma^{3/7} M^{12/7} t^{16/7} \sim \frac{R}{N}
\frac{S^{16/9}}{\lp^2}\ ;
\ee
and the same on the right with $R\rightarrow R_{11}$.
The characteristic Light-Cone scaling in this equation
indicates that indeed the dominant
momentum is set by the charge N on both sides of the diagram.

In the middle and at low temperatures, 
we have phases of smeared boosted eleven dimensional
black holes. The corresponding Gregory-LaFlamme transitions are
complicated by the fact that the momentum is oblique on the two torus.
We qualitatively expect three intermediate phases; a middle one smeared
on two cycles, and two adjacent ones on either sides smeared on a single
cycle. 
There may exist different scaling regimes for
these Gregory-LaFlamme transitions
depending on the size of $\Sigma$. In addition, the torus is skewed
in the near horizon region due to the presence of a non-zero axion
field in the IIB theory. A proper analysis requires a better
understanding of the dynamics of a black hole with an oblique boost on
a skewed two torus. We contend ourselves here with the qualitative picture
just presented.
For higher temperatures, and in the middle of the diagram, 
these complications appear in the form of duality transformations in
IIB theory. The modular group of the torus becomes the SL(2,Z) S-duality
group; the gap between the (N,M) and (-M,N) supergravity vacua is
to be filled with a finite series of such transformations. It appears
from the global structure that
we are guaranteed to find the proper supergravity
framework in every region of this phase. However,
one may need to numerically study
each different combination (N,M) as in~\cite{HASHITZMOR}. None of these
issues will affect the thermodynamics of the middle phase and its
equation of state. We will therefore not dwell in the details of these
duality transformations. 

The final conclusion we are lead to from the lower
part of the diagram is that 1+1 dimensional NCOS theory can describe Light-Cone
M theory with an additional boost. The map between the parameters
of the two theories is given in equations~\pref{lp} and~\pref{radii}.

\begin{figure}
\epsfxsize=10cm \centerline{\leavevmode \epsfbox{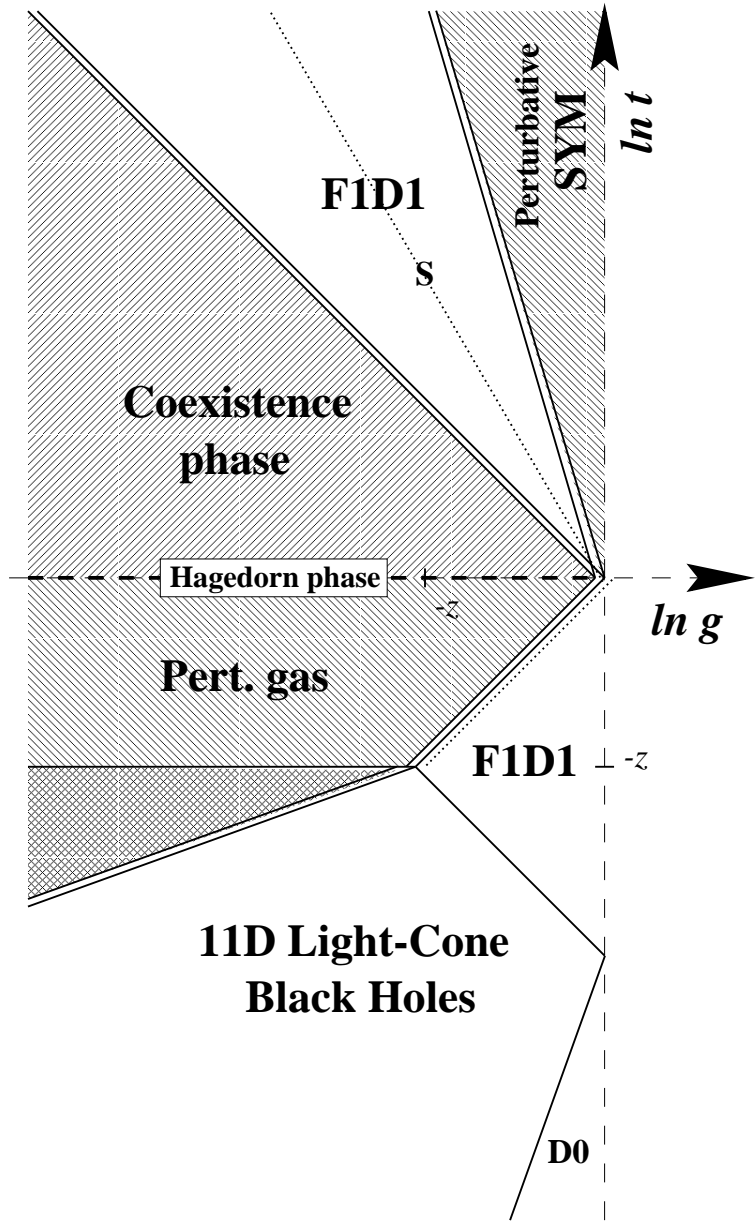}}
\caption{\sl The phase diagram of 1+1 dimensional NCOS theory with one
unit of D string charge.}
\label{fig3}
\end{figure}

Throughout the discussion, the parameter space explored was restricted
to $0\leq z \leq 1$ and $M\gg 1$. Let us next comment one some of the limiting
regimes of these variables. First, we note that the coupling $g$ lives
in a discrete set. For M large, this set is reasonably dense; for example, for
M about a hundred, $\ln g/\ln M$ changes only by a few percents 
for each step near the right border of the diagram; and the distribution
becomes denser towards the left. As M decreases, the right half of the
diagram is progressively less explored by the available values of the
discrete coupling. Drawing the liming case $M=1$ as a separate diagram 
illustrates best this trend. Figure~\ref{fig3} shows the phase diagram
for $M=1$; the axis are $\ln t$ versus $\ln g$.

Let us next vary the size of the circle $z$. For $\Sigma$ in the
range $\le \leq \Sigma \leq \le M$, the structure of the diagram is given 
by Figure~\ref{fig1}. For substringy values of $\Sigma$, the
analysis on the left side of the diagram must be revised. We encounter
a problem at the Gregory-LaFlamme critical point, where the 
transition curve becomes inconsistent with the equation of state
on the other side of the phase. Further to the left, we expect that
the size of the effective degrees of freedom is superstringy; but yet
the cycle $\Sigma$ is substringy. Given these two problems,
it then appears conceptually problematic to extend the thermodynamics
to substringy cycle sizes. Perhaps this indicates that, due to
the non-commutativity of the time-space coordinates, it is inconsistent
or, better said,
dynamically disfavored to compactify the system on a circle smaller
than the length scale set by the non-commutativity.
For larger values of $\Sigma$, the point $\Sigma \sim M\le$
is special, as can be seen from the labels in Figure~\ref{fig4} of
the Appendix. Roughly speaking, the low temperature part and
the right sides of Figure~\ref{fig1}
are scaled out of focus. A point of contention is then the Hagedorn transition,
since we have argued above that it corresponds to the production of
closed strings winding the cycle. We then should expect it to be scaled
out in the large cycle size limit. The phenomenon is also
coupling sensitive as it involves a process where open strings join their
endpoints to form wrapped closed strings. A resolution of
this puzzle is possibly in the following observation:
the region just to the right of the Hagedorn
crossing point, the wedge between the supergravity descriptions, 
cannot be explained through the simplistic analysis we
have presented. This is because the coupling there is becoming big, while the
shape of the wedge we have drawn is an approximate one that breaks down
precisely in this part of the diagram, even within our supergravity analysis. 
We then may expect that
the microscopic dynamics in this area to the left of
$g\sim 1/\sigma$, shaded differently
in Figure~\ref{fig1}, involves dynamics that needs to be understood
by a microscopic calculation. Hence, for $z\rightarrow\infty$,
we propose that
the Hagedorn transition we described above is scaled away towards the left,
and the new unknown dynamics takes over; the latter should obviously not
involve the NCOS closed string sector.
The system exhibits a remarkable level of richness and
more analysis is needed to decode all of its structures.

\subsection{Discussion}\label{disc}

We studied a region of the thermodynamic parameter space of
the 1+1 dimensional NCOS theory on a circle.
Our results can be summarized as follows:

\begin{itemize}
\item We showed evidence that the perturbative expansion of the
NCOS theory is with respect to the parameter $g\equiv G_o\sqrt{M}$.

\item We have found that, at strong coupling, the scale of non-commutativity
gets dressed by a power of the NCOS string coupling.
We get the new characteristic length scale $\le g^{1/3}$.

\item The full SL(2,Z) duality group is needed to explore the
space of relevant supergravity vacua. In particular, we have the
peculiar feature that the dilaton grows at high and low energies
at fixed coupling, reflecting a UV-IR mixing in 
the non-commutative dynamics.

\item Stringy non-commutativity is relevant to the thermodynamics 
at around the energy scale of non-commutativity,
and above it at weak
coupling; away from these regimes, the dynamics appears to simplify.

\item The NCOS theory can describe Light-Cone M theory with an additional
boost. One of the two charges of the system singles out the Light-Cone
direction on a two torus.

\item The Hagedorn transition may indeed involve the transfer of
energy from the open string sector to wrapped closed strings, as suggested
in~\cite{KLEBMALDA}. This is correlated with a peculiar behavior in the
curvature scale of the dual geometry.

\item We find a possible lower bound on the size of the circle of
compactification; mainly, the scale set by the non-commutativity. 
We also note an interesting special point at $\Sigma\sim M \le$.
\end{itemize}

The system we have studied appears to be very rich in structure. 
We have presented a first analysis and a great deal remains to
be explored. The phases appearing on the left of Figure~\ref{fig1}
need to be studied in detail from a microscopic viewpoint. Our
picture of the phase structure in this region may get refined through
a better understanding of the underlying stringy dynamics. 
It would be interesting
to see whether the small cycle size limit can yield new insight into
the theory.
Such an analysis will also lead to a better
understanding of this phase space from the point of view of the SYM
theory. The flipping of the dilaton field in the (N,M) geometry
may be indicative of interesting behavior in the beta function
of the SYM theory. A renormalization group analysis at strong coupling
is possible and may lead interesting results about the dynamics of
the gauge theory in the presence of electric flux~\cite{LINWU}. 
It may also be useful to look at the higher dimensional cases
and OM theory using the same technology.
Finally, understanding
the dynamics of the Gregory-LaFlamme transitions on a skewed two torus
may be an amusing problem from the point of view of gravitational 
dynamics.

In perspective, given the control one has on the perturbative and strong
coupling aspects of this particular stringy system, it appears to be a prime
setting to explore stringy non-commutative physics decoupled from 
complications of a gravitational nature.

\section{The details}\label{details}

In this section, we sketch the technical details involved in constructing
the phase diagram of Figure~\ref{fig1}. There are twelve distinct phases in
the system, eight of which can be described through their supergravity 
duals. We have labeled the various transition curves with small latin letters,
bulk phases with capital letters, and patches of geometries by
a letter and a number. The reader is referred to 
Figure~\ref{fig2} in the Appendix for a roadmap. 
Our starting point is the central phase in Figure~\ref{fig2}. 

\sect{The D1F1 phase (A)}
The IIB theory is parameterized by the string coupling $\gs$ and the
string scale $\alp$. 
The metric of interest is the one found in~\cite{SL2Z}
for the bound state of $N$ fundamental strings and $M$ D strings. In the
string frame, it looks like
\bb\label{mainmetric}
ds_{\str}^2=\gs \lk(\frac{K}{L}\re)^{1/2} \lk[ A^{-1/2} \lk( -f dt^2+
\Sigma^2 dy^2\re)+
A^{1/2} \lk( f^{-1} dr^2+r^2 d\Omega_7^2\re)\re]\ .
\ee
The coordinate $y$ is compact with size $2\pi$.
The NSNS and RR gauge fields are
\bb\label{theothers}
B_{ty}=\gs^2 \Sigma \frac{N}{M} A^{-1} L^{-1/2}\ ,\ \ \ 
A_{ty}=\Sigma \lk(A^{-1}-1\re) L^{-1/2}\ ;
\ee
and the dilaton and axion get turned on as well
\bb
e^{\phi}=\gs A^{1/2} \frac{K}{L}\ ,\ \ \ \chi=\frac{N}{M} A^{-1} \frac{A-1}{K}\ .
\ee
The harmonic functions are
\bb
A=1+\frac{q^6}{r^6}\ ,\ \ \ f=1-\frac{r_0^6}{r^6}\ ,
\ee
with
\bb
q^6\equiv \frac{32 \pi^2 M}{\gs^2} \alp^3 L^{1/2}\ ,
\ee
and $r_0$ the location of the thermodynamic horizon.
The other functions appearing in~\pref{mainmetric} are
\bb
K\equiv 1+ A^{-1} \lk(\frac{N \gs}{M}\re)^2\ ,\ \ \ 
L\equiv 1+ \lk(\frac{N \gs}{M}\re)^2\ .
\ee
Note that we have chosen coordinates such that the S-dual
metric asymptotes to Minkowski space. This corresponds to the choice
advertised in~\pref{strangemetric}. And
the $B_{ty}$ field attains the value~\pref{quant}
at infinity. The coordinate $y$ is compactified on a circle of radius
$\Sigma$. 

The decoupling limit is obtained by 
\bb
\alp\rightarrow 0\ ,\ \ \ \gs\alp=\alpe G_s^2\mbox{     held fixed,}
\ee
while $G_s\rightarrow M/N$. We also need to keep $U\equiv r/\alp$ fixed
\footnote{Note that in the more conventional coordinates 
$r\rightarrow \gs^{1/2} r$, the radial coordinates would scale as
$\sqrt{\alp}$. Our choice corresponds to a canonical normalization
in the S-dual SYM frame instead.}.
In this limit, the metric~\pref{mainmetric} 
becomes (A1)
\bb\label{mainpatch}
ds_{\str}^2=\alp G_s \Delta^{1/2} \lk[
\frac{\le G_s}{\sqrt{32\pi^2 N}} U^3 \lk(-f dt^2+\Sigma^2 dy^2\re)
+\frac{\sqrt{32\pi^2 N}}{\le G_s} U^{-3} 
\lk( f^{-1} dU^2+U^2 d\Omega_7^2\re)\re]
\ .
\ee
The other fields become
\bb
B_{ty}=\alp\Sigma \frac{\alpe^2 G_s^4 U^6}{32\pi^2 N}\ ,\ \ \ 
A_{ty}=\alp^3 \Sigma \frac{G_s U^6}{32\pi^2 N}\ ;
\ee
\bb\label{mainaxion}
e^\phi=\frac{\sqrt{32\pi^2 N}}{\le^3 G_s U^3} \Delta\ ,\ \ \ 
\chi=G_s^{-1} \Delta^{-1}\ .
\ee
The dispersion relation of perturbations propagating in this curved space
dictates a relation between the radial extent $U$ and the energy scale of
the perturbation. The holographic UV-IR relation yields
\bb
U_0^2 \sim \frac{T N^{1/2}}{\le G_s}\sim \frac{M^2}{g^3}\frac{t}{\alpe}\ ,
\ee
where $T$ is the corresponding energy scale or, in our case, the 
Hawking temperature for $U_0$ the location of the horizon. 
The temperature in NCOS string scale units is denoted by
 $t\equiv T \le$, and
$\Delta$ is defined by
\bb
\Delta\equiv 1+ \frac{\alpe^3 G_s^4}{32\pi^2 N} U^6=
1+\frac{g}{32\pi^2} t^3\ ,
\ee
with $g\equiv G_s \sqrt{N}=G_0 \sqrt{M}$ as defined in the Introduction.
$G_0=\sqrt{G_s}$ is the string coupling of the NCOS theory.
There are two temperature regimes in this system, separated by
the scale $t\sim g^{-1/3}\equiv t_c$. For low temperatures, one can
replace $\Delta\rightarrow 1$ in the equations above. While the decoupling
limit is a strict scaling limit, this is an approximation to a 
certain regime lying within the energy window that is survived by the 
limit. We also define
the dimensionless parameter $\sigma\equiv \Sigma/\le$.

The duality boundaries of this vacuum are as follows:

\begin{itemize}
\item The dilaton at the horizon $U=U_0$ becomes big unless (curve (a))
\bb\label{foldeddil}
t\gg M^{-2/3} g \Delta^{2/3}\ .
\ee
For high temperatures $t\gg t_c$, this corresponds to 
\bb
t\ll M^{2/3} g^{-5/3}\ ,
\ee
while for lower temperatures,
\bb
t\gg M^{-2/3} g\ .
\ee
Beyond these points, an element of the $SL(2,Z)$ S-duality group
needs to be applied.

\item The curvature scale of the metric as measured at its horizon 
becomes string scale unless (curve (i))
\bb
t\ll g \Delta\ .
\ee
The high temperature regime yields
\bb
t\gg g^{-1}\ ,
\ee
while the low temperature scenario is
\bb
t\ll g\ .
\ee
Beyond these point, the perturbative 1+1 NCOS theory emerges with coupling
constant given by $g=G_o\sqrt{M}$, string scale $\alpe$, scale of
non-commutativity set by $\alpe$, and $M$ Chan-Paton indices on the
string endpoints. 

\item The size of the circle on which the bound state is wrapped 
becomes string scale, as measured at the horizon, unless (curve (c))
\bb\label{tdual}
t\gg \sigma^{-4/3} g^{-1/3} \Delta^{-1/3}\Rightarrow
t\gg g^{-1/3} \sigma^{-4/3}\ .
\ee
In this case, there is a single temperature regime if $\sigma>1$.
Otherwise, we get $t\gg \sigma^{-2/3} g^{-1/3}$. Many of our subsequent
equations will be sensitive to the assumption that the cycle size
is superstringy or at around the NCOS string scale. We will not discuss
the substringy regime in this work for the reasons discussed in
the Introduction.
Beyond the regime set by equation~\pref{tdual}, 
one needs to look at the T-dual geometry.
\end{itemize}

\begin{figure}[h]
\epsfxsize=12cm \centerline{\leavevmode \epsfbox{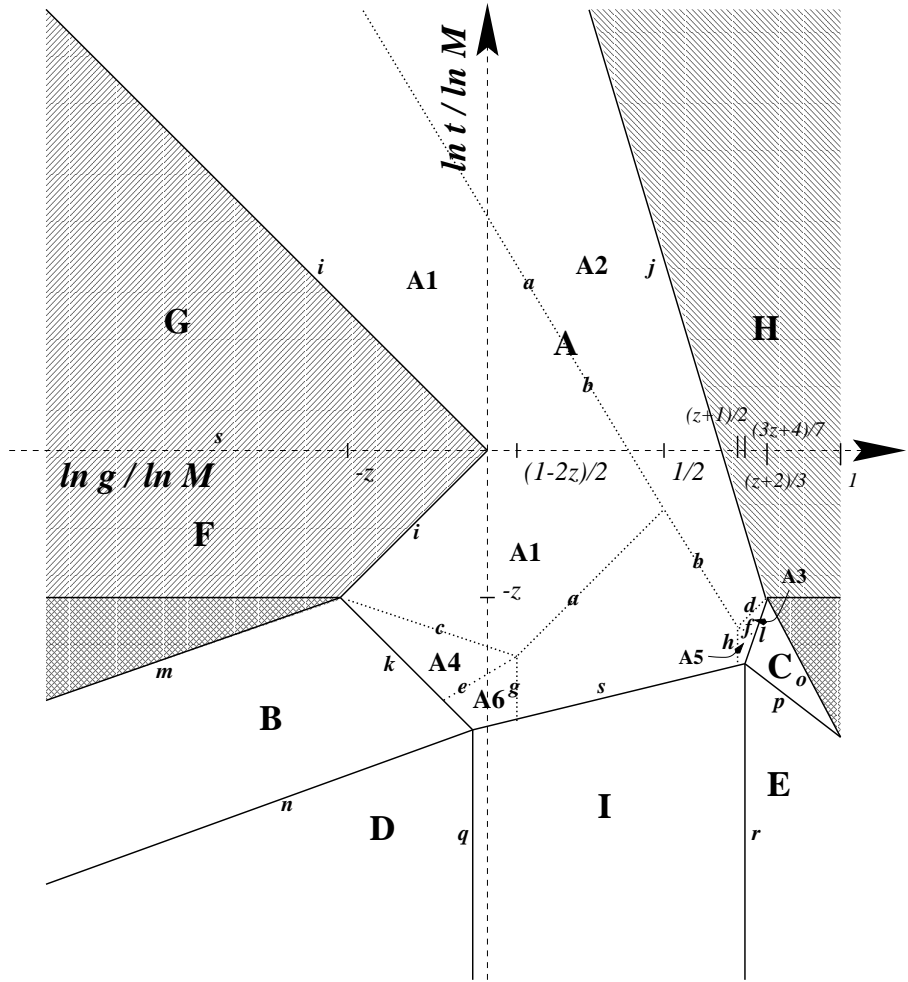}}
\caption{\sl The NCOS phase diagram: roadmap used in section 2.
}
\label{fig4}
\end{figure}

For high temperatures $t\gg t_c$,  we sew onto the strong coupling
phase by applying the S-duality transformation 
\bb
\lk(\begin{array}{c} -M \\ N\end{array}\re)=
\lk( \begin{array}{cc} 0 & -1 \\ 1 & 0 \end{array} \re)
\lk(\begin{array}{c} N \\ M\end{array}\re)
\ee
on the metric~\pref{mainmetric} and~\pref{theothers}. 
This is the vacuum adjacent to
the patch described above, appearing to its right in Figure~\ref{fig4}.
The metric becomes (A2)
\bb
ds_{\str}^2=\alp\lk[ \frac{\le G_s}{\sqrt{32\pi^2 N}} U^3
\lk( - f dt^2 + \Sigma^2 dy^2\re) +\frac{\sqrt{32 \pi^2 N}}{\le G_s} U^{-3}
\lk( f^{-1} dU^2 + U^2 d\Omega_7^2\re)\re]\ ,
\ee
with the NSNS and RR gauge fields exchanged
\bb
{B'}_{ty}=-A_{ty}\ ,\ \ \ {A'}_{ty}=B_{ty}\ .
\ee
The dilaton and axion become
\bb
e^{\phi'}=\frac{\sqrt{32\pi^2 N}}{\le^3 G_s^3 U^3}\ ,\ \ \ 
\chi'=-G_s=-\frac{M}{N}\ .
\ee
Note that the axion field is attracted to a ratio of integers in the
near horizon region. The metric,
the dilaton and the RR gauge field are identical to the case  
dual to $1+1$ SYM without an electric field and an asymptotic constant 
RR gauge field. For $t\ll t_c$, we note that the previous metric,
equation~\pref{mainpatch}, is of similar form; in that case,
the axion field (equation~\pref{mainaxion} with $\Delta\sim 1$), is inverted.

The boundaries of this vacuum are as follows:

\begin{itemize}
\item The dilaton is big at the horizon unless (b)
\bb
t\gg M^{2/3} g^{-5/3}\ .
\ee
Examining this statement in conjunction to~\pref{foldeddil},
reveals that we have a gap between this vacuum configuration
and the previous one for $t\ll t_c$. 
The resolution of this puzzle lies in the realization
that, even though the axion field vanishes at infinity, it has a non-zero
value in the near horizon region, so that the full $SL(2,Z)$ S-duality
group is to fill the gap with the appropriate vacua.
The axion fields on both sides of this gap are ratios of $M$ and $N$;
one being the inverse of the other.
We will come back to this issue at the end of this section.

\item The curvature scale as measured at the horizon is too big in string
units unless (j)
\bb
t\ll M^2 g^{-3}\ .
\ee
Beyond this point, the perturbative 1+1 dimensional $U(N)$ SYM theory emerges 
with $M$ units of electric flux. 

\item The circle parameterized by $y$ is too small in string units as
measured at the horizon unless (d)
\bb
t\gg \sigma^{-4/3} M^{-2/3} g\ .
\ee
For smaller circles, we need to look at the T-dual configuration.
\end{itemize}

We next look at the T-dual of the $(-M,N)$ phase we just discussed.
The IIA metric is given by (A3) (see for example~\cite{BHO})
\bbb\label{athree}
ds_{str}^2 &=& -\alp\lk(\frac{\le G_s U^3}{\sqrt{32\pi^2 N}}\re) f dt^2
+\alp \frac{\sqrt{32\pi^2 N}}{\le G_s \Sigma^2 U^3} 
dy^2 \nonumber \\
&+&\alp \frac{\sqrt{32\pi^2 N}}{\le G_s U^3}\lk(f^{-1} dU^2+U^2 d\Omega_7^2\re)
\ ,
\eee
with the dilaton
\bb
e^\phi=\frac{\lk(32\pi^2 N\re)^{3/4}}{\Sigma \le^{7/2} G_s^{7/2} U^{9/2}}\ .
\ee
There is one form flux due to the axion and the 
two form RR fields of the T-dual
picture
\bb
A_t=\Sigma\ls \frac{\alpe^2 G_s^4 U^6}{32\pi^2 N},\ \ \ 
A_y=-\ls G_s\ .
\ee
This configuration consists of boosted smeared D0 branes.

The new boundaries for this vacuum are:
\begin{itemize}
\item The string coupling is too big at the horizon unless (f)
\bb
t\ll \sigma^{-4/9} M^{2/9} g^{-7/9}\ .
\ee
We then lift to a smeared wave with oblique momentum on a two
torus in an M-theory.

\item The vacuum undergoes the Gregory-LaFlamme transition along $y$ unless (l)
\bb
t\ll \sigma^{-2} M^{-2} g^{-3}\ .
\ee
The new phase beyond this point is described by localized boosted D0 branes.
\end{itemize}

Going to the opposite side of the phase diagram, we next look at the
T-dual of the original $(N,M)$ configuration.
The IIA metric is (A4)
\bbb\label{afour}
ds_{str}^2&=&-\alp \frac{G_s^2 \le U^3}{\Delta^{1/2} \sqrt{32\pi^2 N}}
\lk( \Delta f - \Delta + 1\re) dt^2
-2\frac{\alp}{\Sigma} \frac{\le^3 G_s^2 U^3}{\Delta^{1/2} \sqrt{32\pi^2 N}} 
dt dy \nonumber \\
&+&
\frac{\alp}{\Sigma^2}\frac{\sqrt{32\pi^2 N}}{G_s^2 \Delta^{1/2} \le U^3} dy^2
+\alp \Delta^{1/2} \frac{\sqrt{32 \pi^2 N}}{\le U^3} \lk( f^{-1} dU^2
+U^2 d\Omega_7^2\re)\ ,
\eee
and describes boosted smeared D0 branes again. The coordinate $y$ has period
$2\pi$. The dilaton is
\bb
e^\phi=
\frac{\lk(32\pi^2 N\re)^{3/4} \Delta^{3/4}}{\le^{7/2} G_s^2 U^{9/2} \Sigma}\ ,
\ee
while the one form gauge fields are given by
\bb
A_t=-\ls\Sigma \Delta^{-1} \frac{\alpe^2 G_s^3 U^6}{32\pi^2 N}\ ,\ \ \ 
A_y=\ls G_s^{-1} \Delta^{-1}\ .
\ee

The new boundaries of this patch are:
\begin{itemize}
\item The Gregory-LaFlamme instability along $y$ occurs unless (k)
\bb\label{leftgl}
t\gg \sigma^{-2} g^{-1}\ .
\ee
The new phase consists of boosted localized D0 branes.

\item The dilaton is too big at the horizon unless (e)
\bb
t\gg \sigma^{-4/9} M^{-4/9} g^{5/9}\ .
\ee
Otherwise, we lift to M theory and a configuration of oblique smeared
waves on a torus.
\end{itemize}

We next look at the vacuum 
obtained by lifting the metric~\pref{athree}.
The eleven dimensional metric becomes (A5)
\bbb
ds_{11}^2&=&
\alp \le^{4/3} G_s^{4/3} \Sigma^{2/3} \lk( f^{-1} dU^2 + U^2 d\Omega_7^2\re)
\nonumber \\
&+&\alp \frac{\le^{4/3} G_s^{4/3}}{\Sigma^{4/3} } dx^2
-\alp f \frac{\le^{10/3} G_s^{10/3} \Sigma^{2/3} U^6}{32\pi^2 N} dt^2
\nonumber \\
&+&\frac{ 32 \pi^2 \alp}{\le^{14/3} G_s^{14/3} \Sigma^{4/3} N U^6} 
\lk( N dx_{11} - M dy + \frac{\le^4 G_s^4 \Sigma U^6}{32\pi^2} dt \re)^2\ .
\eee
$x_{11}$ is compact with periodicity $2\pi$, so that the combination
$N x_{11} - M y$ respects the periodicity of the torus.
Note also that we have chosen a somewhat unconventional normalization
such that the M theory energy scale is scaled by $\gs^{1/3}$; \ie\ 
we have lifted to M theory using a dilaton field which asymptotes to $\gs$.
This is convenient in this setting since it
makes the $\alp$ scaling of this region of space explicit.

There are two new boundaries to this phase:
\begin{itemize}
\item A reduction to IIA theory along $x$ is needed unless (h)
\bb
g\gg M^{1/2} \sigma^{1/2} \ .
\ee
We come back to this issue at the end of the section.

\item A localization transition occurs unless  (s)
\bb\label{midloc}
t\gg \sigma^{-1/2} M^{-1} g^{1/2}\ .
\ee
The new phase is a boosted black hole in M theory. 
More about this phase below.
\end{itemize}

Jumping to the other side of the phase diagram, we next describe the
M lift of the T dual of the $(N,M)$ phase.
The geometry is again that of smeared waves 
on a two torus in an M theory (A6)
\bbb
ds_{11}^2&=&
\alp \frac{\le^{4/3}}{\Sigma^{4/3} G_s^{2/3}} 
\lk( dy- \frac{\le^4 G_s^4 U^6 \Sigma}{32\pi^2 N} dt\re)^2 \nonumber \\
&-&\alp\frac{\le^{10/3} G_s^{10/3} \Sigma^{2/3} U^6}{32\pi^2 N} f dt^2
+\alp \le^{4/3} G_s^{4/3} \Sigma^{2/3} \lk( f^{-1} dU^2+U^2 d\Omega_7^2\re)
\nonumber \\
&+&\alp \frac{32\pi^2}{\le^{14/3} G_s^{11/3} \Sigma^{4/3} M U^6}
\lk( M dx_{11} + N dy - \alpe^2 \frac{\Sigma G_s^4 U^6}{32\pi^2}dt\re)^2\ ,
\eee
with the periodicities of $x_{11}$ and $y$ being $2\pi$.

The boundaries are:
\begin{itemize}
\item We need to perform a reduction to IIA theory unless (g)
\bb
g\ll \sigma^{-1} M^{1/2}\ .
\ee
The emerging vacuum is.

\item And a localization transition occurs unless (s)
\bb
t\gg \sigma^{-1/2} M^{-1} g^{1/2}\ ,
\ee
as in~\pref{midloc}.
The emerging phase is a boosted black hole in M theory.
\end{itemize}

We now have enough structure to describe all the boundaries of the
D1-F1 phase. Strictly speaking however, 
to complete the discussion about all the supergravity vacua patching up
this phase, we need to write down the metrics appearing in the gap
region created in the middle of the phase. 
These various frames appearing in the
middle patch are to be obtained by making use of the full $SL(2,Z)$
S-duality group. The
fact that $M$ and $N$ are relatively primed is to play an important
role in the existence of the appropriate duality transformation. 
We refrain from going into a detailed discussion since this will not
affect any of the physical results of the work, mainly the phase 
transitions of the system.

The duality transformations sewing all these vacua into the D1F1 patch
do not change the equation of state of this phase. The latter is given
by
\bb\label{maineos}
\varepsilon \equiv E \le \sim \sigma M^2 g^{-1} t^3\ .
\ee
This is identical to the equation of state obeyed by a system of $M$
D1 branes. This point was emphasized in~\cite{MALDARUSSO} in the context of
the D3 brane system with magnetic and electric fluxes.

\sect{The boosted D0 phase (B)}
This phase arises from the metric~\pref{afour} via
a Gregory-LaFlamme transition along $y$.
The equation of state is given by
\bb\label{boosteos}
\varepsilon \sim \sigma^{3/5} M^2 g^{-6/5} t^{14/5}\ .
\ee
To obtain this equation, we boost localized D0 branes
\bb\label{disp}
E^2-p^2= \lk( M_0+\mu \re)^2 \Rightarrow E\simeq p
+\frac{M_0^2}{2 p } + \frac{M_0 \mu}{p}\ ,
\ee
where $M_0$ is the BPS mass of $M$ D0 branes, $p$ is the
boost momentum, and $\mu$ is the excitation energy above extremality.
The limit follows in the scaling regime under consideration, \ie\ 
the infinite boost scenario, since
\bb
M_0=\frac{M \Sigma}{\alp}\rightarrow \infty \ ,\ \ \ 
p=\frac{N\Sigma}{\alp^2}\gs \alp\rightarrow \infty \ ,
\ee
obtained from the BPS masses of M fundamental strings and N D-strings
\footnote{We remind the reader that the canonical normalization of the
energy we use is in the $(-M,N)$ frame; \ie\ the metric asymptotes
to Minkowski in this frame.}. We then use the equation of state of
D0 branes, being careful to substitute the appropriate asymptotic
values of of the fields in the relevant frame
\bb
\mu\sim T S\sim \frac{\lk(\gs \alp\re)^{1/6}}{\ls \Sigma^{1/3}} 
S^{14/9} M^{-7/9}
\times \lk(\gs \re)^{1/2}\ .
\ee
The last factor of $\gs^{1/2}$ arises from the choice of coordinates in
the metric~\pref{mainmetric}. Substituting these into~\pref{disp} yields the
equation of state~\pref{boosteos}. To check for consistency of our analysis,
we equate equation~\pref{boosteos} with~\pref{maineos} to find the boundary 
transition curve
\bb
t\sim \sigma^{-2} g^{-1}
\ee
which is exactly~\pref{leftgl}, determined there independently 
from the shape of the geometry.

There are two new boundaries to this phase.
Instead of looking for them in complicated geometries, 
to find these phase transition curves we will use a trick:
\begin{itemize}
\item From~\cite{MSSYM123}, we know that localized D0 branes
at rest undergo two phase transitions; either when the entropy $S$ is of
$M$, the number of D0 branes (a Gregory-Laflamme transition to
a boosted black hole); or when $S\sim M^2$, which is the point
where the curvature scale at the horizon becomes string scale.
These statements, making reference to the density of states and the
number of D0 branes must be boost invariant. Using $E\sim T S$ in
the equation of state~\pref{boosteos}, and $S\sim M^2$, we find the transition 
curve (m)
\bb
t\sim \sigma^{-1/3} g^{2/3}
\ee
in terms of the temperature of the system. Beyond this point, the 
perturbative NCOS may emerge. 

\item Similarly, a localization transition will occur at $S\sim M$, or (n)
\bb
t\sim \sigma^{-1/3} M^{-5/9} g^{2/3}\ .
\ee
Beyond this point, we have the phase of an eleven dimensional localized
black hole boosted obliquely along a torus. 
\end{itemize}

\sect{The phase of D0 branes (C)}
The equation of state is given by
\bb
\varepsilon\sim g^{-2/5} M^{8/5} \sigma^{3/5} t^{14/5}\ .
\ee

Using the same trick as in the previous subsection, we get the transition
curves:
\begin{itemize}
\item $S\sim N^2=M^4/g^4$ for the string scale curvature point (o)
\bb
t\sim \sigma^{-1/3} M^{4/3} g^{-2}\ .
\ee
Beyond this temperature, we have a coexistence phase leading up to
perturbative $1+1$ SYM gas.

\item $S\sim N$ corresponds to the localization transition into a
boosted black hole, which translates to the temperature (p)
\bb
t\sim \sigma^{-1/3} M^{2/9} g^{-8/9}\ .
\ee
\end{itemize}

\sect{The boosted 11D black holes (D and E)}
These are phases of localized black holes boosted obliquely on a two torus.
The equation of state was given in~\pref{bbheos}; along with the
parameters of the corresponding M theories in equations~\pref{lp} 
and~\pref{radii}. On curves of the qualitative form (q) and (r)
in Figure~\ref{fig4}, these phases undergo Gregory-LaFlamme transitions
to smeared and boosted black holes.

\sect{The smeared boosted black holes (I)}
The middle phases at low temperatures consist of various smeared 
and boosted black holes. The complications arising in this phase 
were discussed in the Introduction. We have not determined the
exact scaling of the phase transitions curves separating these
phases amongst themselves and from the rest of the phase diagram.
There presence is required for the consistency of the structure of
the diagram.

\sect{Perturbative 1+1 SYM and coexistence phase (H)}
The perturbative SYM phase obeys, to leading order in the Yang-Mills coupling,
the equation of state
\bb
\varepsilon \sim \sigma N^2 t^2\ .
\ee
Its effective coupling becomes of order one one both curves bounding
it on the diagram. The lower one is also associated with finite
size effects. 

\sect{Perturbative 1+1 NCOS and coexistence phase (F and G)}
These phases can be understood through a thermodynamic
interplay between the open and closed string sectors of the NCOS theory.
The details were discussed in the Introduction. 

\paragraph{\bf Acknowledgments:}
I thank P. Argyres, M. Moriconi, S. Minwalla for discussions; and
S. Apikyan and the Yerevan State University for hospitality.
This work was supported by NSF grant 9513717.

\newpage
\section{Appendix}

\newpage
\providecommand{\href}[2]{#2}\begingroup\raggedright\endgroup

%\bibliography{biblio}
%\bibliographystyle{utphys}

\end{document}